\documentclass[aps, prb, twocolumn, a4paper, superscriptaddress]{revtex4-2}
\usepackage{graphicx} 
\usepackage{dcolumn} 
\usepackage{amsmath, bm} 
\usepackage{xr-hyper}
\usepackage{hyperref} 
\usepackage{mathrsfs} 
\usepackage{color} 
\usepackage{gensymb} 
\usepackage{braket}
\usepackage[utf8]{inputenc}
\usepackage{tikz}
\usepackage{lipsum}
\usepackage{makecell}
\usepackage{multirow}
\DeclareMathOperator\im{\textrm{i}}
\usetikzlibrary{positioning}
\usetikzlibrary{shapes,arrows,arrows,positioning,fit}

\makeatletter

\newcommand{\vecc}[1]{\ensuremath{\boldsymbol{#1}}}
\newcommand{\kp}{$\vecc{k}$-point}

\newcommand{\vecr}{\ensuremath{\boldsymbol{r}}}

\newcommand{\laplace}{\nabla^2}
\newcommand{\castep}{\textsc{Castep}}

\newcommand{\vecdiv}{\vecc{\nabla} \cdot}

\def\sto{Sm$_2$Ti$_2$O$_7$}
\def\nzo{Nd$_2$Zr$_2$O$_7$}
\def\dto{Dy$_2$Ti$_2$O$_7$}
\makeatother

\begin{document}

\title{First-principles calculations of magnetic states in pyrochlores using a source-corrected exchange and correlation functional}

\author{Z.~Hawkhead}
\affiliation{Department of Physics, Centre for Materials Physics, Durham University, Durham DH1 3LE, United Kingdom}

\author{N.~Gidopoulos}
\affiliation{Department of Physics, Centre for Materials Physics, Durham University, Durham DH1 3LE, United Kingdom}
\author{S.~J.~Blundell}
\affiliation{Oxford University Department of Physics, Clarendon Laboratory, Parks Road, Oxford OX1 3PU, United Kingdom}
\author{S.~J.~Clark}
\affiliation{Department of Physics, Centre for Materials Physics, Durham University, Durham DH1 3LE, United Kingdom}
\author{T.~Lancaster}
\affiliation{Department of Physics, Centre for Materials Physics, Durham University, Durham DH1 3LE, United Kingdom}

\begin{abstract}
We present a first-principles investigation of the spin-ice state in \dto\ using a magnetic source-free exchange and correlation functional, implemented in the {\sc Castep} electronic-structure code.
By comparing results from the conventional local spin-density approximation, we show that a spin-ice state in \dto\ can be reliably obtained  by removing the magnetic sources from the exchange and correlation contributions to the potential, and we contrast this against the computed ground states of other frustrated pyrochlore magnets. 
\end{abstract}
\maketitle



\section{Introduction}

Materials that show  long-range, non-collinear magnetic spin textures occur widely in Nature. 
A system with collinear order
has a global quantisation axis along which all the spins are aligned or antialigned, while
for a non-collinear state, each ordered spin can potentially have a different local direction. Such non-collinear states are commonly found, for example, in systems where the spin-orbit interaction leads, in a low-energy approximation, to single-ion anisotropy or to a Dzyaloshinski-Moriya interaction.

Following the initial incorporation of collinear magnetism into density functional theory (DFT) within the local density approximation, attempts were made to describe non-collinear magnetism using the formalism of DFT.
Starting in the 1980s, Kubler {\it et al.}, \cite{kubler1988density}  made progress in describing non-collinear spins via a self-consistent method. 
To incorporate spin-orbit coupling required the development of a fully relativistic treatment \cite{Yamagami}, in which all electrons are treated non-collinearly within the spin-polarised coupled Dirac equation.
After development of the methods, the main hindrance in the use of DFT  for non-collinear magnetism is the accuracy of the functionals. 
The optimised effective potential (OEP) method \cite{PhysRev.90.317, PhysRevA.14.36, hollins2012optimized} has been shown \cite{OEP} to successfully
describe the  magnitudes of  magnetic moments, and  therefore represents a candidate route to extend DFT's exchange and correlation (xc) functionals. 
Sharma {\it et al.} \cite{sharma2007first} extended the spin-density formalism to non-collinear spin via the OEP, which has the advantage that it does not rely on the local collinearity of spins required when applying standard  functionals.
Eich and Gross \cite{eich2013transverse} made a similar extension  
within a local density-like approximation.

Non-collinearity arises in a number of magnetic compounds that adopt the pyrochlore lattice, a three-dimensional (3D) arrangement of corner-sharing tetrahedra, which is known to exhibit a high degree of geometrical frustration. 
For example, the 
pyrochlore oxide Dy$_2$Ti$_2$O$_7$ \cite{Blote1969} exhibits an interesting non-collinear magnetic structure owing to the strong Ising-like crystal-field anisotropy at each Dy site.
The Dy$^{3+}$ ions are magnetic and the $J =15/2$ manifold is split by the crystal field, leading to a $\approx$10$\mu_{\rm B}$ ground state moment.  The ground state is separated from crystal-field excited states by a gap of a few hundred Kelvin \cite{Bertin2012}. The crystal-field anisotropy constrains the magnetic moments to lie along the local $\langle 111 \rangle$ axes \cite{harris1997geometrical,Harris1998}. 
An effective ferromagnetic coupling between these moments results from the combination
\cite{DenHertog2000} of dominant long-range dipolar interactions ($D = $1.41~K) with antiferromagnetic nearest-neighbour exchange ($J = -3.72$~K) \cite{Castelnovo2011}. 
As a result of this combination of interactions and the local anisotropy, below about 1~K the system settles into a disordered spin-ice state.  
This state is characterized by a `2-in 2-out' spin configuration (meaning that two spins point in  and two spins point out of each tetrahedron), analogous to proton displacement vectors in Pauling’s model of hydrogen disorder in water ice, the
residual configurational entropy measured for these materials being close to Pauling’s predicted value for ice \cite{Ramirez1999,bramwell2001spin}.
The excitations in spin ice are created by reversing a single spin, which thereby produces a pair of magnetic monopoles which can move independently through the lattice 
\cite{castelnovo2008magnetic}.  

Studying these monopole excitations has become of great current interest 
\cite{morris2009dirac,Dusad2019,hallen2022}, but an important scientific aim is to understand the local electronic properties if \dto\ in more detail, particularly as the monopole transport may arise from the precise local arrangement of spins \cite{Tomasello2019,hallen2022}.  Hitherto there have  been limited first-principles simulations of the spin-ice state in \dto, and the spin-ice physics is generally understood within the low-energy description of the underlying magnetic energy-levels described above, invoking only crystal field levels and single-ion anisotropy effects. 
The bulk of the first principles work carried out on pyrochlore materials focuses on electronic and structural effects. Early work investigating the electronic structure of a range of pyrochlores, including \dto, had success matching x-ray emission spectra~\cite{nemoshkalenko2001electronic}.  Similar success has also been achieved in using DFT to study further pyrochlore materials~\cite{hinojosa2008first,kang2014effect}. DFT has also been applied to compare the results of neutron scattering experiments in magnetic pyrochlores by calculating the phonon spectra~\cite{ruminy2016first}.  Little work has been done from first principles on the magnetic configurations of \dto, although there have been  attempts at understanding magnetic behaviours in other magnetic pyrochlore materials~\cite{amirabbasi2020ab,huebsch2022magnetic,iqbal2017signatures}.

Since the ordered spin-ice magnetic structure is inherently non-collinear and highly degenerate, it provides a challenge for spin-DFT calculations, particularly with the xc functionals currently at our disposal. Here we present results of spin-DFT calculations where we make use of a source-corrected version of the well-known local spin-density approximation (LSDA) functional to stabilise the spin-ice state. 
By making comparisons to calculations performed using the conventional LSDA, we  show that realising the spin ice state is  possible because of the use of the source-corrected functional. 
We also present calculations on two further pyrochlore materials, \nzo\ and \sto, which are known to host an ordered \textit{all-in, all-out} (AIAO) spin texture~\cite{mauws2018dipolar,bertin2015nd,xu2015magnetic}.
These latter  examples show that the source-free LSDA correctly predicts that the AIAO state is of lower energy than a spin-ice state, which is not stable in the two materials considered.

\section{Source-free methods}
\subsection{Theory}
The  natural description of magnetism using electronic structure uses
the spin density, or equivalently the magnetisation, 
which is a continuous vector field that  is discretised onto a grid in order to perform calculations.
For calculations of magnetic properties there are then two key parameters: the charge density $n(\vecr)$ and the magnetisation density $\vecc{m}(\vecr)$.
The accuracy of a DFT calculation depends  on the choice of xc functional \cite{lehtola2018recent,sousa2007general}, with 
different  functionals capturing different aspects of a physical system.
In standard DFT we have an xc functional $E_\mathrm{xc}[n]$ and an associated xc-potential, $V_\mathrm{xc}(\vecr)$, given by the functional derivative:
\begin{equation}\label{Eq:V_xc}
  V_{\mathrm{xc}}(\vecr)=\frac{\delta E_\mathrm{xc}[n(\vecr)]}{\delta n(\vecr)}.
\end{equation}
In calculating the xc spin-potential for a non-collinear density 
we cannot use the analogue of  Eq.~\ref{Eq:V_xc}, since the density in a non-collinear treatment is not a scalar field.
Instead the electron potentials can be expressed as 2$\times$2 matrices, where we use spinors to account for the vector magnetisation~\cite{gidopoulos2007potential}. We can express the xc spin potential in terms of a non-spin potential and a magnetic field via a four-component potential \cite{gidopoulos2007potential,sharma2018source,capelle2001nonuniqueness}
\begin{equation}
  \mathcal{V}_\mathrm{xc}(\vecr) =  V_{\mathrm{xc}}(\vecr)\mathrm{I}_{2} +  \mu_\mathrm{B} \vecc{B}_\mathrm{xc}(\vecr) \cdot \vecc{\sigma},
  \label{Eq:spin_pot}
\end{equation} 
where  the magnetic field $\vecc{B}_\mathrm{xc}$ is the vector part of the xc potential, $\vecc{\sigma}$ is the vector of Pauli spin matrices, $\mu_\mathrm{B}$ is the Bohr magneton, and $\mathrm{I}_{2}$ is the $2\times 2$  identity matrix.
We can then relate each term in Eq.~\ref{Eq:spin_pot} separately to $E_{\mathrm{xc}}$: (i) the scalar potential $V_\mathrm{xc}(\vecr)$ is found using Eq.~\ref{Eq:V_xc}, taking the density to be the scalar part of the non-collinear density;
(ii) the vector term, or the magnetic field, is given by
\begin{equation}\label{Eq:V_xc_ncm}
 \vecc{B}_\mathrm{xc}(\vecr)=-\frac{\delta E_\mathrm{xc}[n(\vecr),\vecc{m}(\vecr)]}{\delta \vecc{m}(\vecr)}.
\end{equation}

Much research has gone into developing xc functionals specific to non-collinear spin~\cite{sharma2007first,eich2013transverse}.
However, there is no  widely-used, accurate functional that consistently replicates experimentally-observed  magnetic states, and 
most currently available xc functionals are simple extensions to functionals designed for collinear systems. 
To make use of functionals  such as the  LSDA, at each point in space we  rotate the vector spin-density such that it lies along the $z$-axis, allowing us to decompose it into spin-up ($n_\uparrow$) and spin-down ($n_\downarrow$) densities which can be used to calculate $E_\mathrm{xc}$~\cite{kubler1988density,eschrig2001density}. 
We then take the functional derivative of $E_{\mathrm{xc}}[n_\uparrow, n_\downarrow]$ with respect to $n_{\uparrow}$ and $n_{\downarrow}$. This yields $\vecc{B}_\mathrm{xc}$ along the $z$-axis.
Then, we carry out the inverse rotation on $\vecc{B}_\mathrm{xc}$ and we finally obtain a non-collinear vector $\vecc{B}_\mathrm{xc}$, which however remains locally parallel (at every point in real space) to the magnetisation density, or the spin density.
As the resulting energy is calculated point-wise, the method imposes a non-physical constraint that the magnetisation must be locally collinear with $\vecc{B}_\mathrm{xc}(\vecr)$.

\citet{sharma2018source} highlight another important problem with standard functionals.
From the Maxwell equations, for any arbitrary magnetic field, $\vecc{B}$, the divergence of the field should be zero ($\vecc{\nabla} \cdot \vecc{B}= 0$),
which follows from the absence of magnetic sources.
However, this condition is not met for the common functionals, LSDA and PBE, which give results consistent with magnetic sources existing on the surfaces of a sample of the material. \citet{sharma2018source} suggest a method for improving these functionals that we reproduce in outline here.
Starting from the Helmholtz theorem, a vector field can be decomposed into two components: one of which is divergence free and one that is curl free~\cite{Riley:345425}, 
\begin{equation}\label{Eq:helmholtz}
\vecc{B}(\vecr)=\vecc{\nabla}\times\vecc{A} + \vecc{\nabla}\phi.
\end{equation}
To ensure that the magnetic field is source free we must explicitly subtract the term $\vecc{\nabla}\phi$ that contributes to the divergence of the field.
We have the freedom to select the gauge and chose $\phi$ such that it is the solution to the Poisson equation,
\begin{equation}\label{Eq:poisson}
\nabla^2\phi(\vecr) = -4\pi\vecc{\nabla}\cdot\vecc{B}.
\end{equation}
The source-free magnetic field 
$\tilde{\vecc{B}}$ can then be constructed using
\begin{equation}\label{Eq:b_sf}
\tilde{\vecc{B}}(\vecr)\equiv \vecc{B}(\vecr) +\frac{1}{4\pi}\nabla\phi(\vecr),
\end{equation}
and we can then obtain the xc magnetic field in terms of the substitution,
\begin{equation}
  \vecc{B}_\mathrm{xc}(\vecr) \rightarrow s\tilde{\vecc{B}}_{\mathrm{xc}}(\vecr),
\label{eqn:forp}
\end{equation}
where $s$ is an empirical scaling parameter.


We have implemented this source-free method in the plane-wave pseudopotential code \castep.
The source-free LSDA described by \citet{sharma2018source} is an alteration of the existing functional.
Each time the xc energy and potential is calculated, the potential is used to construct the xc magnetic field $\vecc{B}_\mathrm{xc}(\vecr)$.
Using the procedure described above, we can then calculate the source-free field $\tilde{\vecc{B}}_\mathrm{xc}(\vecr)$ and reconstruct the spin-potential.
It is this potential that is then used in the Hamiltonian of the system.
Critically, this is not a one-shot approach: the correction is calculated every time the xc potential is required and is therefore self-consistent.


To implement the method, we make use of the plane-wave basis set in order to efficiently solve the Poisson equation in Eq.~\ref{Eq:poisson}.
In a plane-wave basis, $\phi(\vecr)$ can be expressed as 
\begin{equation}
\phi(\vecr)=\sum_{\vecc{G}_j} c_{\vecc{G}_j}e^{\im\vecc{G}_j\cdot\vecr},
\label{Eq:phi_expansion}
\end{equation}
where $\vecc{G}_{j}$ are the reciprocal lattice vectors and $c_{\vecc{G}_{j}}$ are the Fourier expansion coefficients of $\phi(\vecr)$.
The Fourier expansion allows us to efficiently compute the Laplacian of the scalar potential $\phi$, 
\begin{equation}\label{Eq:laplace_phi}
\laplace\phi(\vecr)=-\sum_{\vecc{G}_j}|\vecc{G}_j|^2 c_{\vecc{G}_j} e^{\im\vecc{G}_j\cdot\vecr}.
\end{equation}
We can also express $\vecc{B}_\mathrm{xc}(\vecr)$ in terms of its Fourier coefficients $\vecc{b}_{\vecc{G}_j}$:
\begin{equation}\label{Eq:Bxc_coeff}
\vecc{B}_\mathrm{xc}(\vecr) = \sum_{\vecc{G}_j}\vecc{b}_{\vecc{G}_j}e^{\im \vecc{G}_j\cdot\vecr}.
\end{equation}
For a given vector $\vecc{G}_j$, the Fourier coefficients of $\vecdiv \vecc{B}_\mathrm{xc}(\vecr)$ follow from
\begin{equation}\label{Eq:B_coeff}
  \mathscr{F}[\vecdiv \vecc{B}_\mathrm{xc}(\vecr)](\vecc{G}_j)=\im\vecc{G}_j\cdot\vecc{b}_{\vecc{G}_j},
\end{equation}
where $\mathscr{F}$ denotes the FT.
By rearranging Eq.~\ref{Eq:poisson} and substituting the reciprocal space expressions in Eq.~\ref{Eq:laplace_phi} and Eq.~\ref{Eq:B_coeff}, we come to an expression for the Fourier coefficients of $\phi(\vecr)$,
\begin{equation}
c_{\vecc{G}_j}=\frac{\im\vecc{G}_j\cdot\vecc{b}_{\vecc{G}_j}}{4\pi|\vecc{G}_j|^2}.
\label{phi_solved}
\end{equation}
Knowing these coefficients, we can build the source-free magnetic field and reconstruct the spin-potential using Eq.~\ref{Eq:spin_pot}.
We note that the corrected xc potential is no longer strictly local as we use knowledge of the gradient of the potential when solving the Poisson equation. 


\subsection{Tests on elemental magnetic materials}
To test the validity of the implementation of the source-corrected LSDA, we performed calculations on a range of elemental magnets and magnetic compounds.
The main focus of the testing is on body-centred cubic (bcc) or $\alpha$-Fe, where we performed magnetic calculations on a geometry optimised unit cell.
Total energies are converged to better than 10~meV using a 7$\times$7$\times$7
Monkhorst Pack 
(MP) \kp-grid and a plane-wave cut off radius of 1600~eV.
We included spin-orbit coupling (SOC) and used norm-conserving relativistic pseudopotentials throughout. 
Similarly, other calculations are converged to better than 100~meV/atom.
Each of the materials tested have a ferromagnetic phase which we investigate using a quantisation axis aligned with the crystallographic $c$-direction.
We initialised spin along this direction in each material to ensure that the energy minimisation returned a state with long-range magnetic order.

A useful way to assess the correction to $\vecc{B}_\mathrm{xc}$ is to visualize the magnetic field lines, since
it is easy to demonstrate that the source terms have been  removed from the field.
\begin{figure}[!h]
  \centering
  \includegraphics[width=1\linewidth]{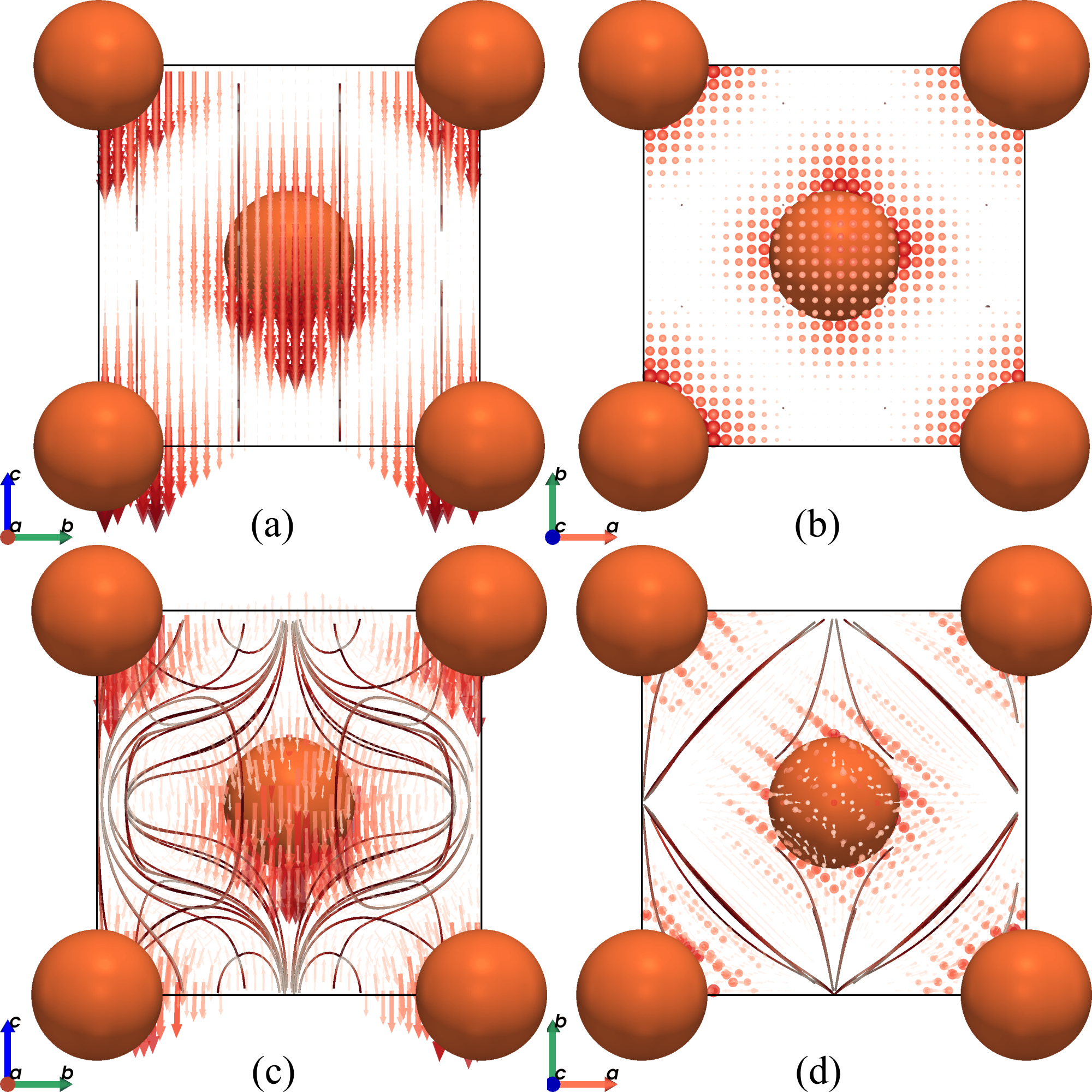}
  \caption{Field lines  for $\mathbf{B}_{\mathrm{ex}}$ in $\alpha$-Fe calculated with the source-corrected and source-uncorrected LSDA. (a,b) xc field calculated using the conventional LSDA functional. (c,d) xc fields calculated using the source-corrected LSDA functional. Colour of the field lines represents the magnitude of the field at that point, from white at low fields to red at high fields. Vectors show the direction of the field flow with size and colour representing magnitude. Where the field magnitude is lower than 10\% of the maximum, vectors have been omitted for clarity. Magnetic moments lie in the negative $c$-direction in both cases. }
  \label{fig:Fe_lines}
\end{figure}
In Fig.~\ref{fig:Fe_lines}, we compare the field lines due to $\vecc{B}_\mathrm{xc}$ in $\alpha$-Fe.
Using both the conventional LSDA [Fig.~\ref{fig:Fe_lines}(a,b) and the source-free LSDA [Fig.~\ref{fig:Fe_lines}(c,d)] we find magnetic moments on each atom aligned along the $c$-axis.
For the LSDA [Fig.~\ref{fig:Fe_lines}(a,b)], the field lines are parallel throughout the entire infinite crystal, implying the presence of a magnetic source at the surface of the system.
It is clear that these field lines for $\alpha$-Fe are also globally collinear with the magnetisation, which is one of the  issues with the LSDA highlighted by \citet{sharma2018source}.

If instead we study the field lines due to the source-corrected functional [Fig.~\ref{fig:Fe_lines}(c,d)] we see that they  form closed loops around the Fe ions.
As we have used the same quantisation axis in both calculation, the magnetisation is also collinear with the $c$-axis when using the source-free LSDA.
However, it is clear then that the field lines for $\vecc{B}_\mathrm{xc}$ are no longer constrained to be locally collinear with the magnetisation when the source terms are removed.

\begin{figure}[!h]
  \centering
    \includegraphics[width=\linewidth]{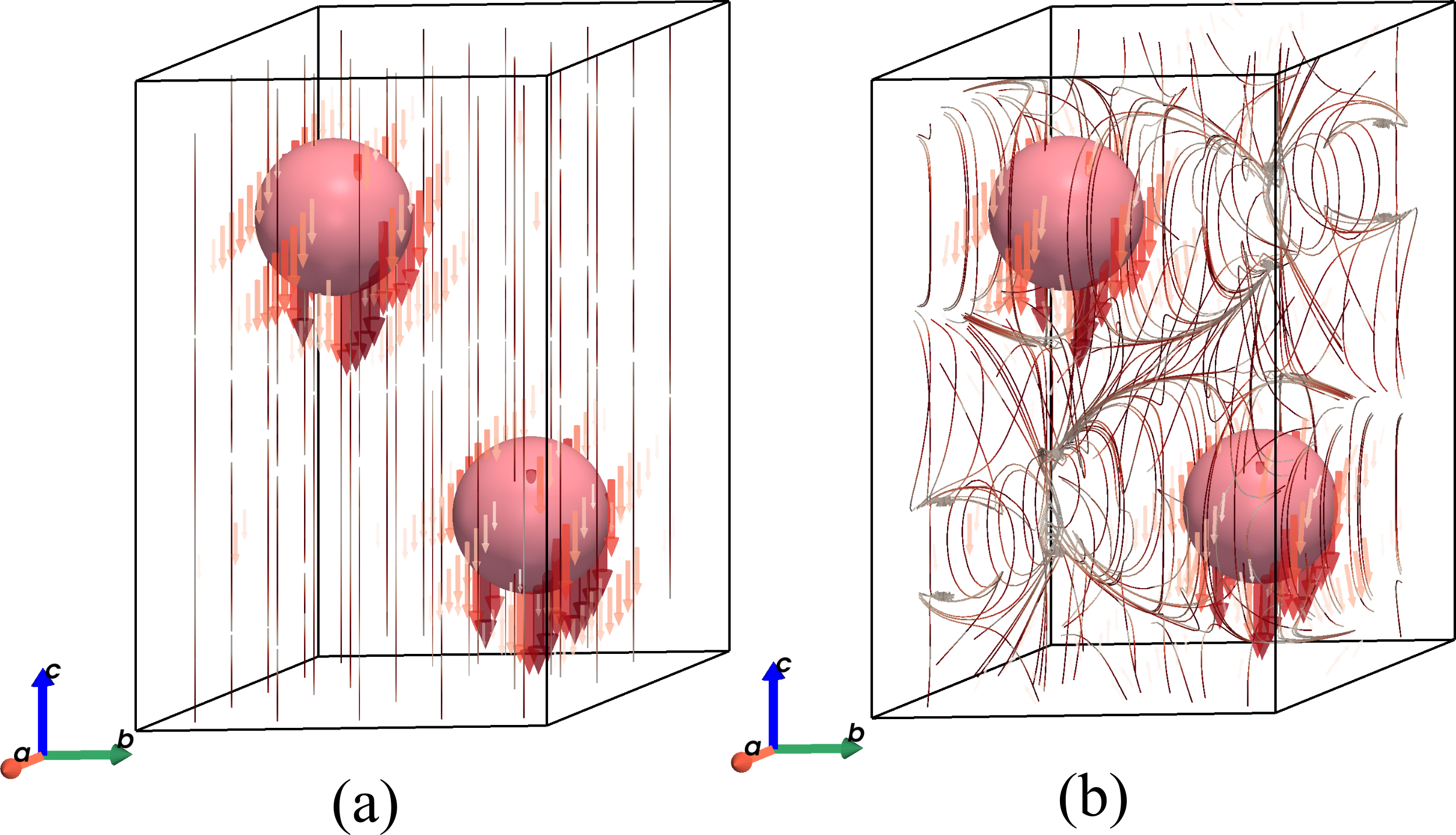}
  \caption{Magnetic field lines in hcp-Co calculated with (a) conventional LSDA and source-free LSDA.  Colour of the field lines represents the magnitude of the field at that point, from white at low fields to red at high fields. }
  \label{fig:hcp-Co}
\end{figure}

We see similar results for the $\vecc{B}_\mathrm{xc}$ field lines in 
hexagonal close-packed (hcp) Co (Fig.~\ref{fig:hcp-Co}).
Using the LSDA [Fig.~\ref{fig:hcp-Co}(a)] we again see parallel field lines aligning with the magnetisation which lies along the $c$-axis for both Co ions.
From the source-corrected LSDA [Fig.~\ref{fig:hcp-Co}(b)] we see the non-collinearity of the field lines.
Close to the Co ions, field lines emerge from the centre of the ion and terminate again at the centre of the ion.
The behaviour further from the Co in the interstitial region is more complicated. 

An example of the effects of the source-corrected functional on the xc field lines on a non-elemental magnet, FeTe, is shown in the Supplemental Material \cite{sm}.
In this case, the field lines flow between the layers of Fe ions past the Te ions.
It is less obvious that the xc field lines in FeTe represent a source-free field.
We lose the simplicity of the elemental magnets by including non-magnetic ions which complicate the exchange interactions.
However, we still see the improved non-collinearity arising from the source-free functional.

\section{Spin ice ground states}

We now turn to the application of the source-corrected methodology to the problem of the spin-ice magnetic structures. 
In order to model a spin-ice state in \dto\,
we  used a primitive unit cell including 22 atoms, with lattice parameters of 7.19~\AA\ calculated by DFT structural relaxation~\cite{osti_1206767}.
The spin structure was calculated with a 5$\times$5$\times$5 MP \kp-grid with a plane-wave cut off of 840~eV.
Convergence testing for \dto\ is shown in the SM \cite{sm}.
For  calculations on \nzo\ and \sto\ we used plane-wave cut-off energies of 843~eV and 860~eV respectively with a 5$\times$5$\times$5 MP \kp-grid grid.
The SCF calculation was performed using the ensemble density functional theory (EDFT) minimisation scheme.
We performed identical calculations treating xc with both conventional LSDA and the newly-implemented source-free LSDA to compare the resulting spin configurations.
In the case of both conventional LSDA and the source-corrected LSDA we initialise a non-collinear spin on each of the Dy ions along the direction of the local $\langle111\rangle$ direction in the spin ice configuration.
Based on testing of the scaling parameter $s$ in the SM~\cite{sm},
we conclude that there is no \textit{a priori} reason to use a any value other than $s = 1$ in Eq.~\ref{eqn:forp}.


\begin{figure}[!h]
  \centering
  \includegraphics[width=7cm]{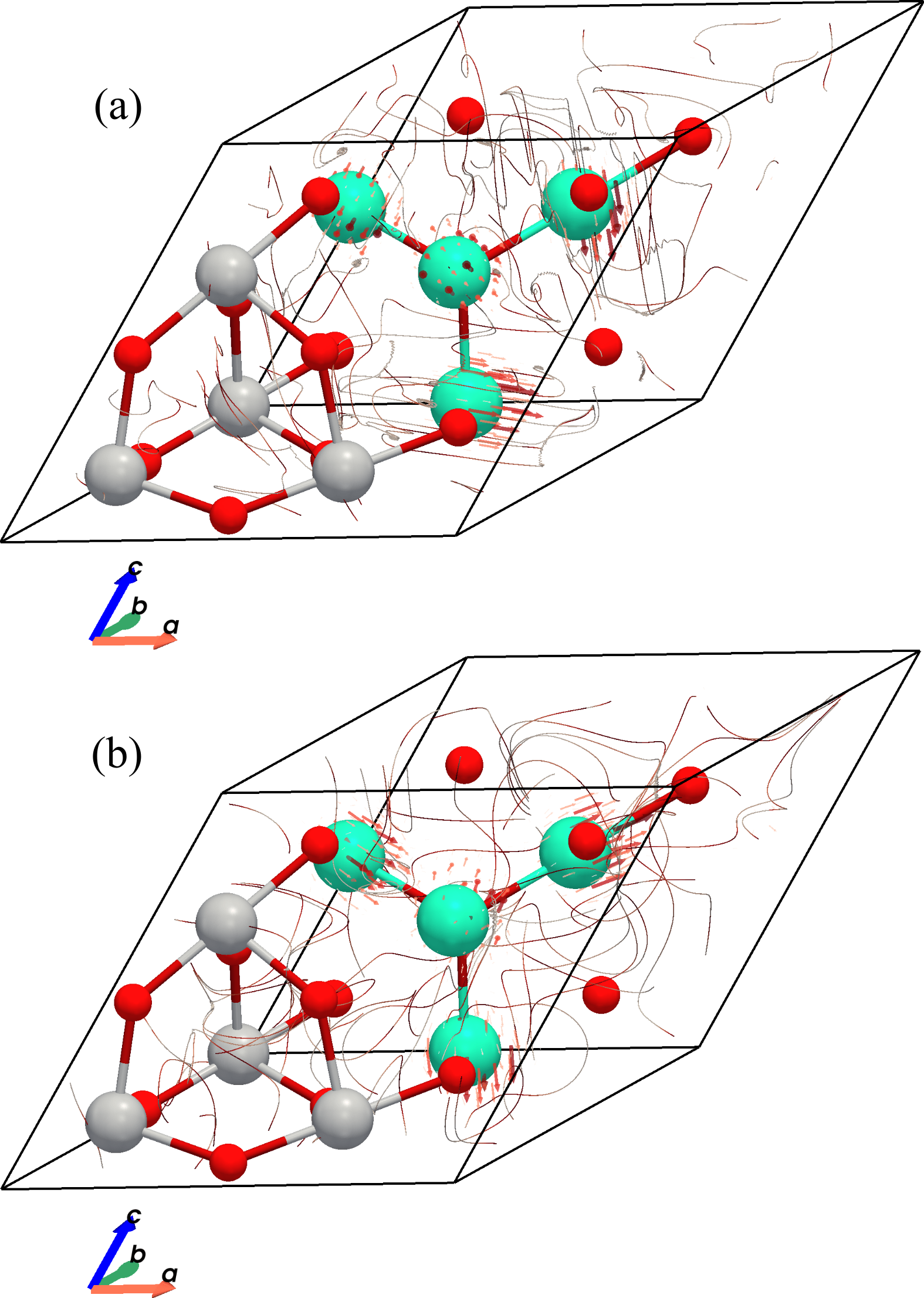}
  \caption{Magnetic field lines of $\mathbf{B}_\mathrm{xc}$ for a primitive unit cell of \dto\ calculated using the (a) conventional LSDA and  (b) source-corrected LSDA, The opacity of the field lines represents the relative strength of the field.  For the conventional LSDA field there is a local collinearity with the spin projected onto the Dy atoms, whereas for the source free functional $B_\mathrm{xc}$ is no longer aligned with the magnetisation. Dy, Ti and O ions are shown in green, silver and red respectively. }
  \label{fig:DyTiO_ice_lines}
\end{figure}

We first examine the magnetic field lines of $\boldsymbol{B}_{\mathrm{xc}}$  in \dto.
The xc field lines for both functionals are shown in Fig.~\ref{fig:DyTiO_ice_lines}, where 
 it is still possible to see that these remain locally collinear around the Dy ions in the case of the conventional LSDA [Fig.~\ref{fig:DyTiO_ice_lines}(a)], aligning with the magnetisation which is localised around these ions.
In the interatomic regions it is less clear that the field lines are collinear, largely due to the lack of significant spin density. Instead these regions are dominated by numerical noise.
However, for the source-free  functional [Fig.~\ref{fig:DyTiO_ice_lines}(b)], the field lines display more obvious non-collinearity and no longer  follow the magnetisation. 
  We  show below that by better capturing the physics of the internal fields using the source-corrected LSDA, we are able to realise the observed ground-state magnetic structure.
\begin{figure}[!t]
  \centering
    \includegraphics[width=\linewidth]{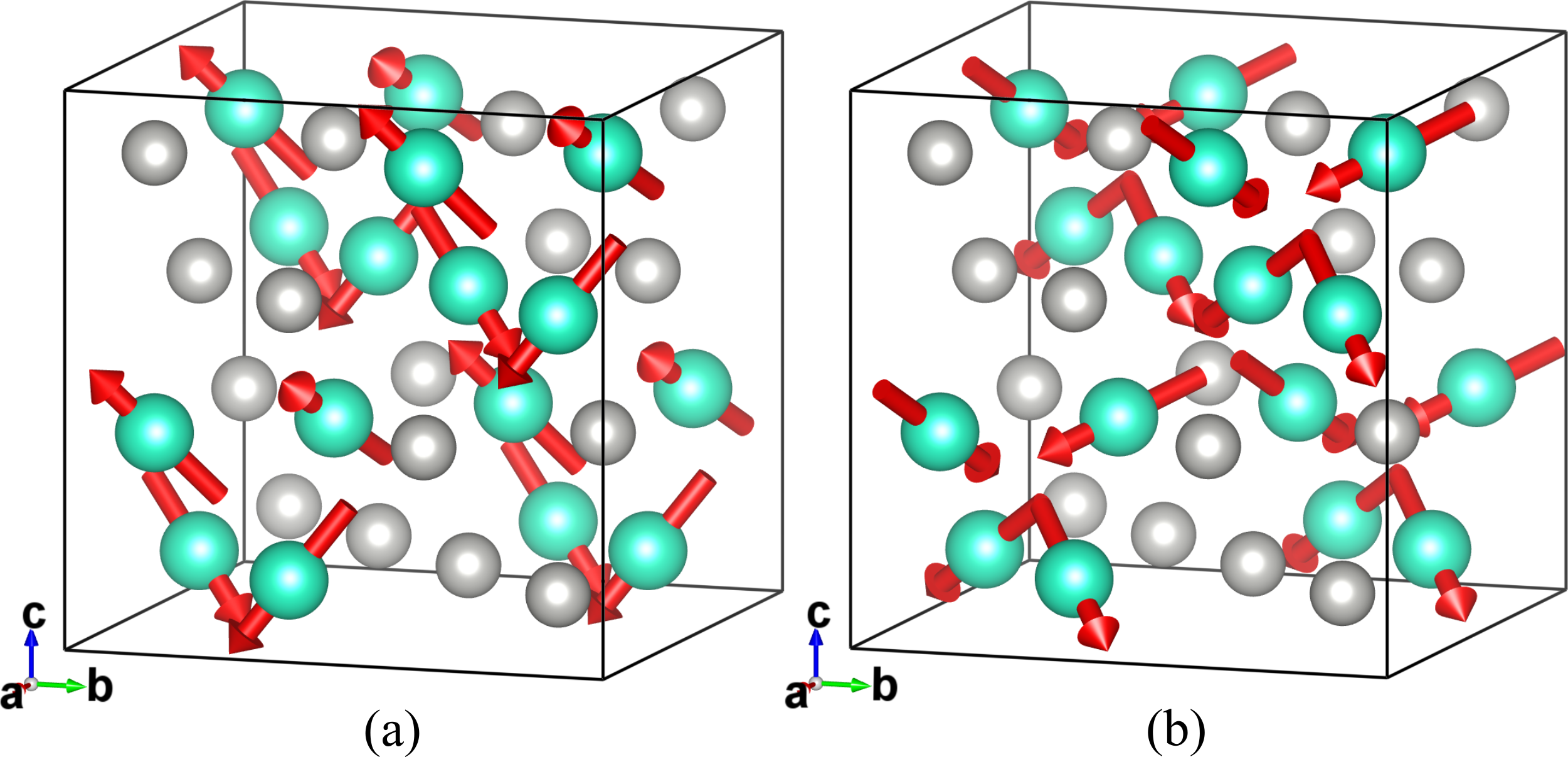}
  \caption{Example magnetic configurations of \dto\ calculated using (a) conventional LSDA, and (b) source-free LSDA, shown in the conventional unit cell. Dy ions and Ti ions are shown in green and silver respectively. The red arrows show the non-collinear spin density projected onto a local atomic basis for the Dy ions. Oxygen atoms are not shown. }
  \label{fig:DyTiO_ice}
\end{figure}

The improvement in non-collinearity provided by the source-free LSDA leads us to realise a spin-ice structure in \dto\ which we find is not possible using conventional LSDA, despite using the same spin-ice initialisation.
The resulting spin structures are shown in Fig.~\ref{fig:DyTiO_ice} where we have taken our spin density and projected it onto the Dy ions using Mulliken analysis.
These calculations were performed a number of times, and in each case for the given convergence parameters, we realise a spin-ice state using the source-free LSDA functional.
Conversely, while conventional LSDA results in a non-collinear configuration of spins, there is no well-defined magnetic structure and a different arrangement is found each time [e.g.\ Fig.~\ref{fig:DyTiO_ice}(a)],
indicative of the randomly-initialised orbitals falling into a different local minimum each time we perform the calculation.
By removing the unphysical source terms from the $\boldsymbol{B}_\mathrm{xc}$ it appears we improve the energy landscape which aids in the minimisation, and
the source-free functional is then able to reliably reproduce the  spin-ice structure in \dto\ [Fig.~\ref{fig:DyTiO_ice}(b)].
As was seen with the magnetic moment of our set of test materials (see Ref.~\cite{sm}), the magnetic moment on the Dy$^{3+}$ is increased under the source-free functional, from 5.0$\mu_\mathrm{B}$ using conventional  LSDA to 5.2$\mu_\mathrm{B}$ with the source free functional. We note that these are significantly smaller than the ordered-moment sizes seen in experiment. Although this could have motivated a different choice of the scaling parameter $s$, we did not do this and set $s=1$ throughout.

\begin{figure}[!t]
  \centering
    \includegraphics[width=5.5cm]{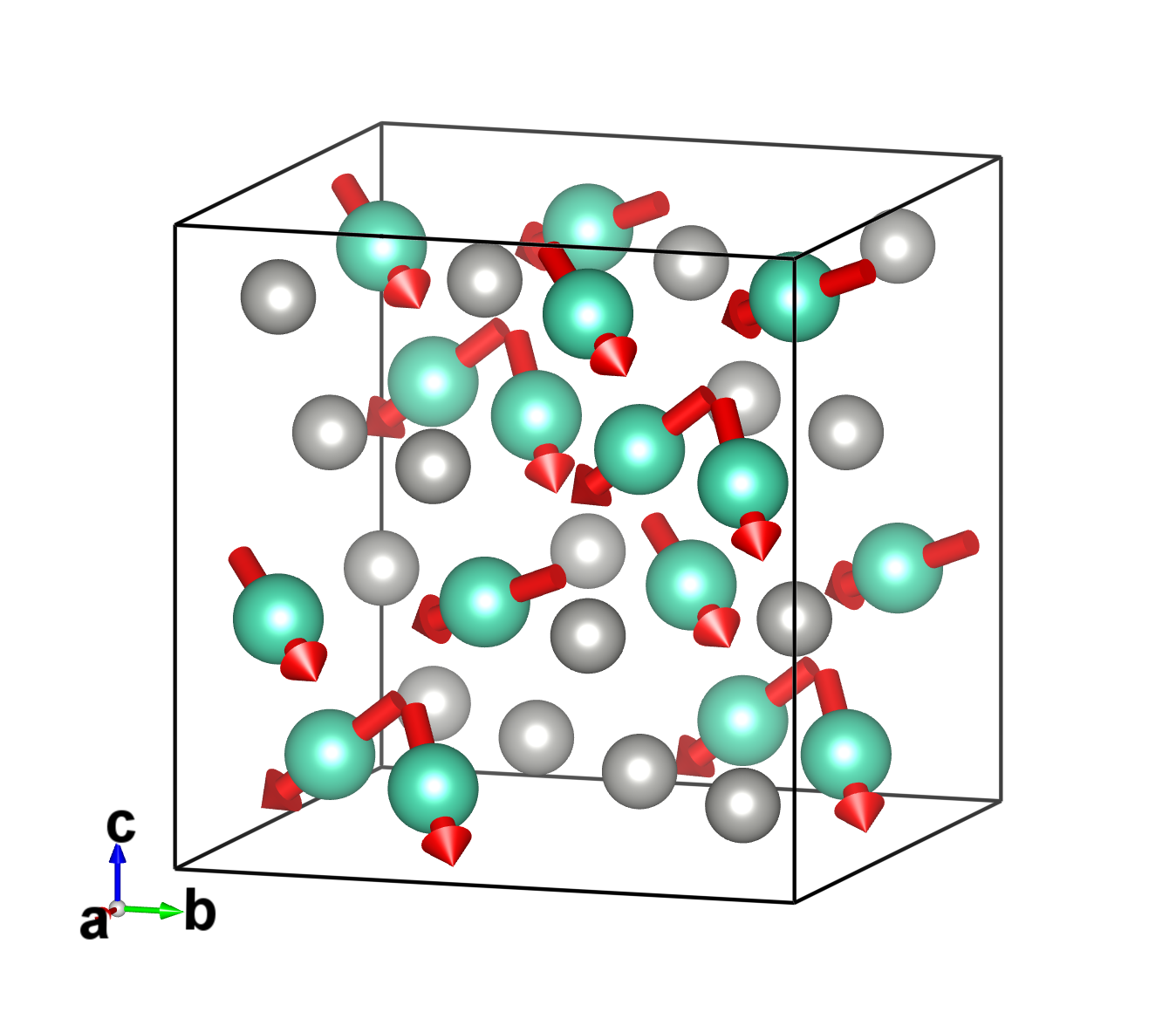}
  \caption{Example magnetic configurations of \dto\ calculated using an all-in, all-out spin initialisation. Dy ions and Ti ions are shown in green and silver respectively. The red arrows show the non-collinear spin density projected onto a local atomic basis for the Dy ions. Oxygen atoms are not shown.} 
  \label{fig:Dy2Ti2O7_aiao_conv}
\end{figure}

To further test the capabilities of the source-free functional, we performed  calculations on \dto\ with the initial spin in an AIAO configuration. 
We see in Fig.~\ref{fig:Dy2Ti2O7_aiao_conv} that the final spin orientation has not remained in the AIAO state, instead it has begun to fall into the spin-ice state seen above, and correspondingly
 find that the total energy of the spin-ice state in Fig.~\ref{fig:DyTiO_ice}(b) is lower in energy than the all-in, all-out calculation.

\begin{figure}[!t]
  \centering
    \includegraphics[width=4.cm]{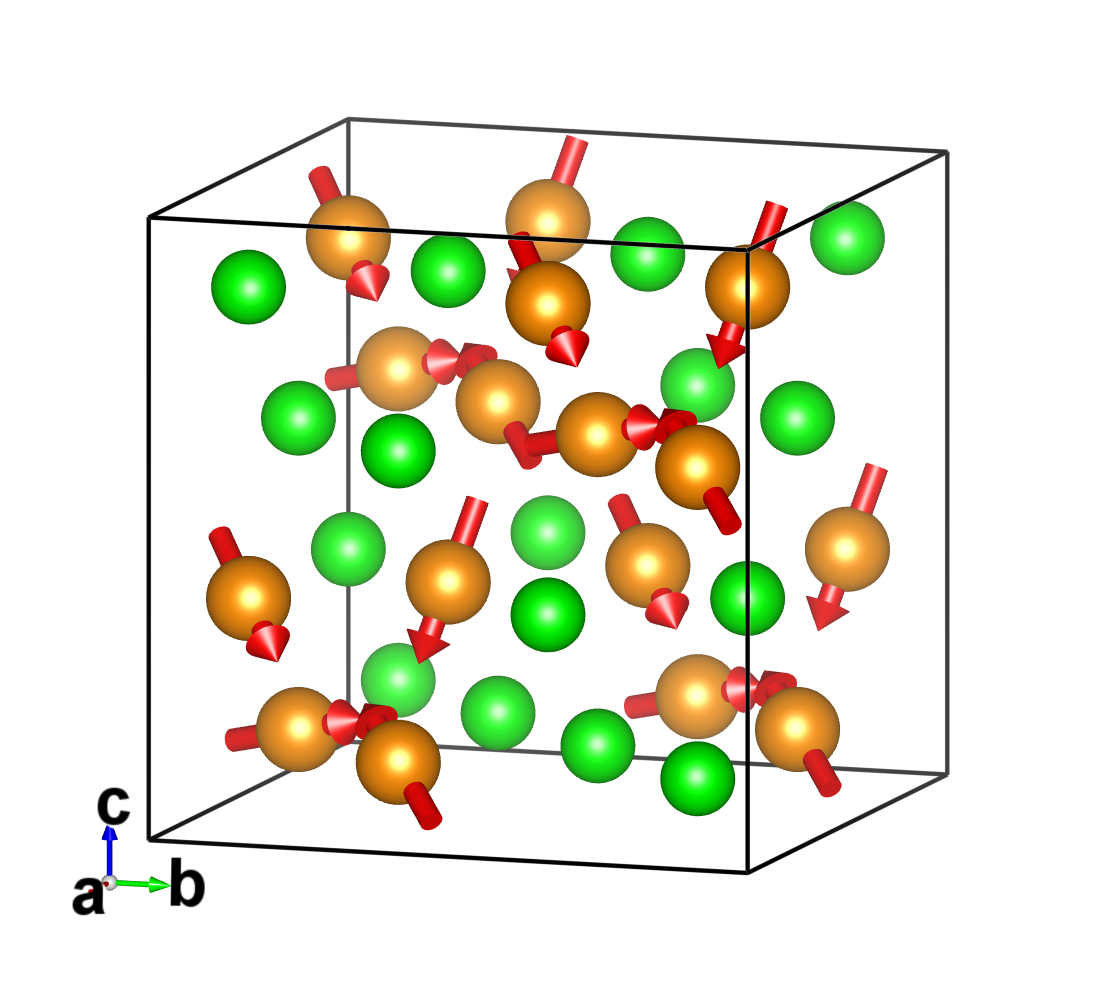}
    \includegraphics[width=4.cm]{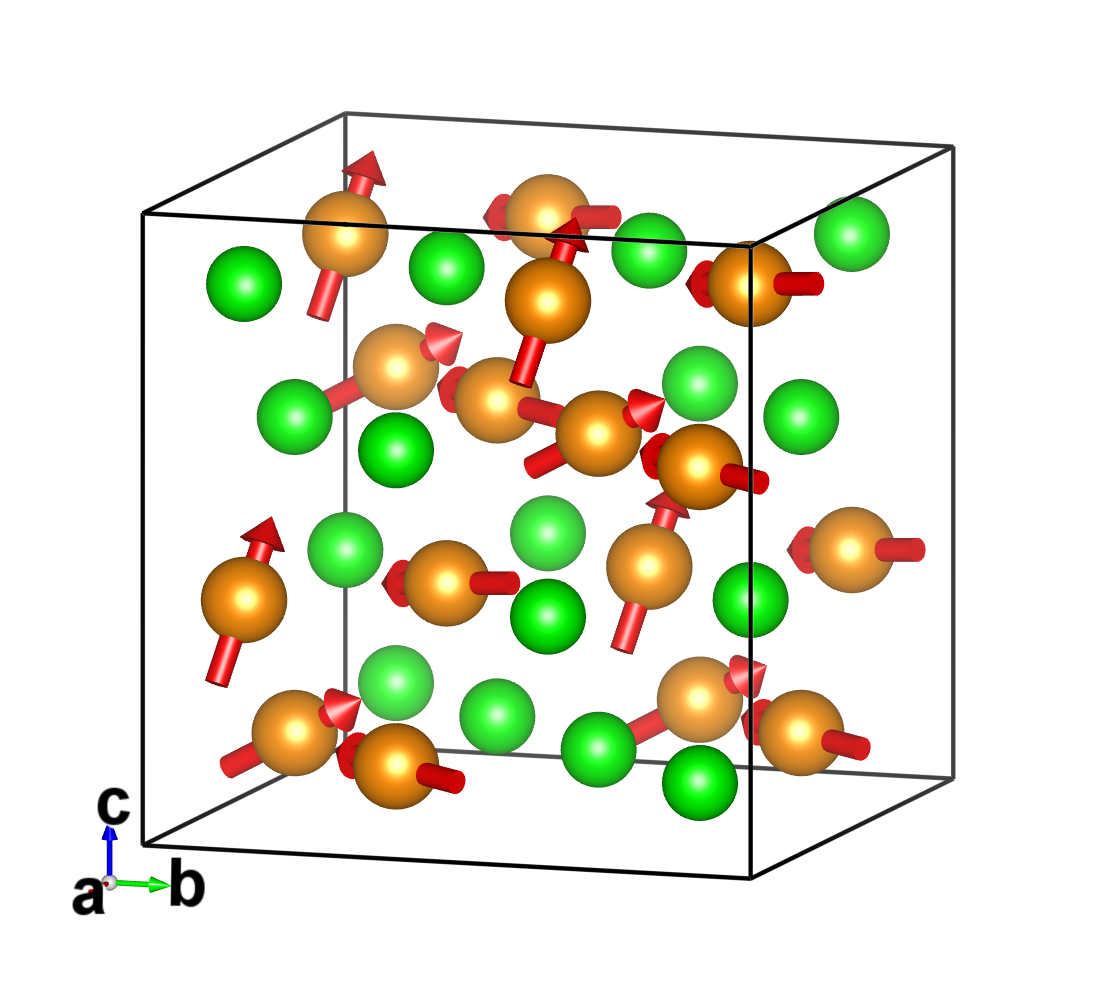}
    \includegraphics[width=4cm]{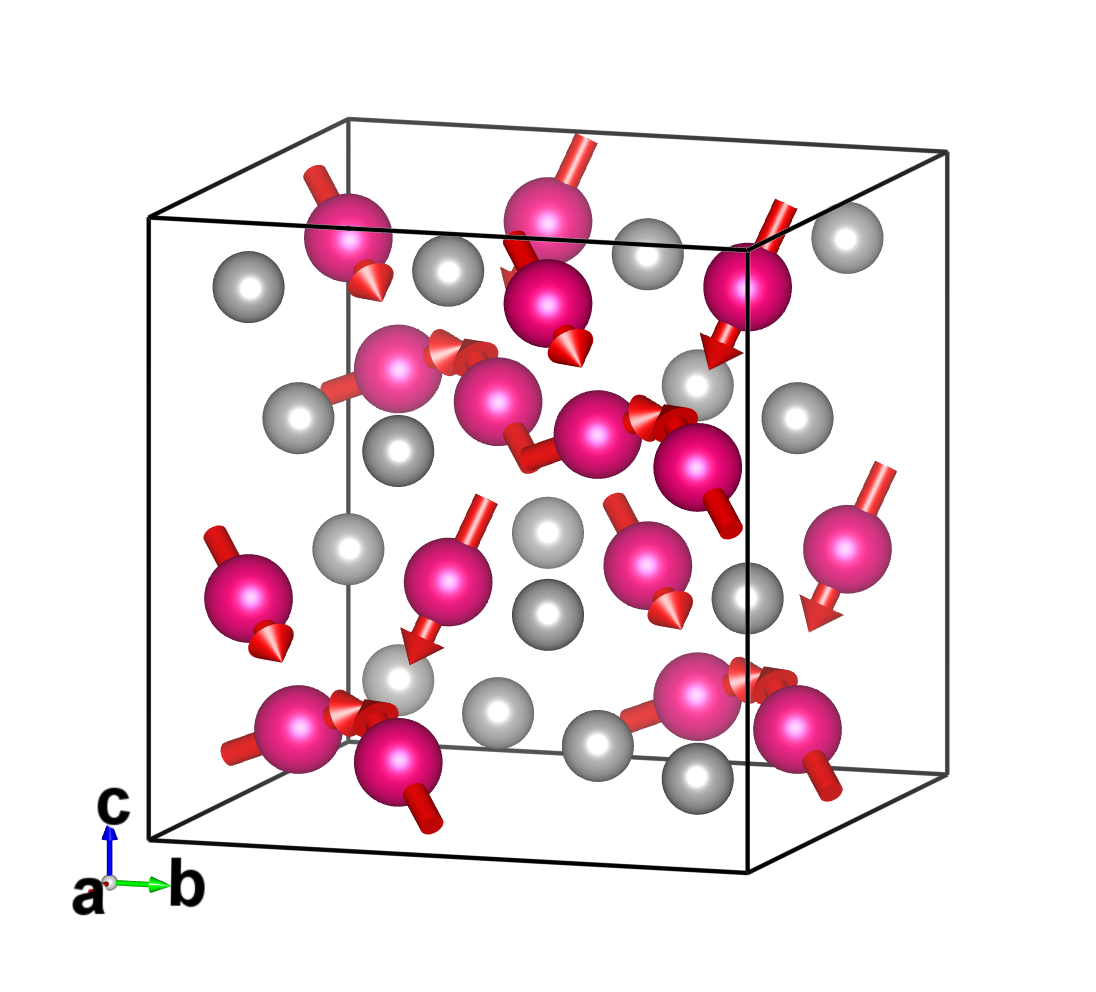}
    \includegraphics[width=4cm]{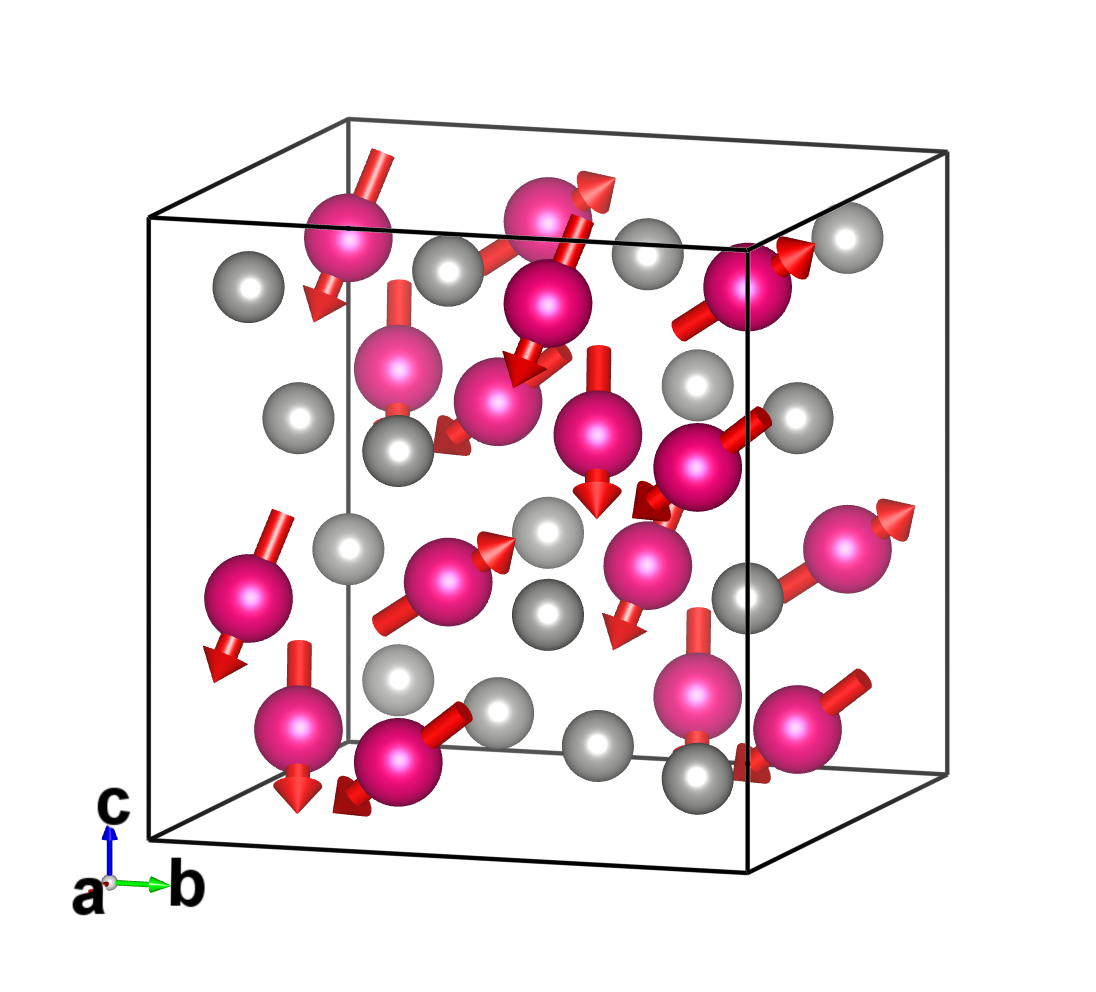}
  \caption{(Top) Example magnetic configurations of \nzo\ calculated using two different spin initialisations, (left) all-in, all-out and (right) spin-ice configurations. Nd ions and Zr ions are shown in orange and green respectively. The red arrows show the non-collinear spin density projected onto a local atomic basis for the Nd ions. 
  (Bottom) Example magnetic configurations for \sto\ calculated using two different spin initialisations, (left) all-in, all-out and (right) spin-ice configurations. Sm ions and Ti ions are shown in pink and silver respectively.
  Oxygen atoms are not shown.} 
  \label{fig:Nd2Zr2O7_aiao_conv}
\end{figure}

To show that the source-free functional provides systematic improvement to the magnetism in the wider class of materials, we have also performed calculations on \nzo\ and \sto\ which are known to have an AIAO ground-state magnetic structure~\cite{lhotel2015fluctuations,xu2015magnetic,mauws2018dipolar}.
We initialised each calculation with both a spin-ice state and an AIAO state. 
In the case of \nzo, the final spin configurations are shown in Fig.~\ref{fig:Nd2Zr2O7_aiao_conv}(top).
For the AIAO initialisation, we find that the resultant spin state is very similar to the initialisation, with some deviation from the local $\langle111\rangle$ direction. 
However, when we compare it to the result of a computation following initialisation in the spin-ice structure, which is known not  to be a stable state in this material, we see a random arrangement of spins. 
This is much like the conventional LSDA calculations of \dto\ shown in Fig.~\ref{fig:DyTiO_ice}(a).
The final AIAO state is found to be 0.4~eV lower in energy than the state found by initialising a spin-ice. 
We see similar results for \sto\  [Fig.~\ref{fig:Nd2Zr2O7_aiao_conv}(bottom) where we return an AIAO state and we fail to find a spin-ice configuration.
In this latter case, the total energy for the AIAO state is found to be  1.0~eV lower that for state found following  spin-ice initialisation. 
Interestingly, by initialising the spin-ice state in \sto, the final state is found to be similar to the $\psi_2$ ground state seen in a different pyrochlore, Er$_2$Ti$_2$O$_7$~\cite{lago2005magnetic,mcclarty2009energetic,de2012magnetic}.

%

\section{Conclusions}

In conclusion, we have implemented a recently-developed xc functional in a plane-wave code \castep, which provides a correction to the LSDA that removes magnetic sources from the resulting xc magnetic field, with the aim of improving the ability to describe non-collinear magnetic spin states.
We have demonstrated that our implementation of this functional  by performing calculations on a number of simple magnetic materials.   
We applied the functional to the famous spin-ice material \dto\ and
find that where the LSDA is unable to capture the spin-ice state when initialised in this configuration, the source-free functional reproducible realises a spin-ice configuration.
We hope that the availability of the source-free xc functional in \castep\ might allow the calculation of exotic magnetic textures which were previously inaccessible to DFT.

\section{Acknowledgments}
This work used the ARCHER UK National Supercomputing Service (http://www.archer.ac.uk) and is supported by EPSRC (UK) (EP/P022782/1).
We are grateful for computational support from Durham Hamilton HPC.
This work is supported by EPSRC (UK) [EP/N032128/1]. 
Research data will be hosted at XXX.

\bibliography{bib}


\end{document}